\documentstyle[a4,12pt]{article}
\begin{document}

\input {psfig.tex}

\vspace*{5mm}

\begin{center}
{\LARGE  \bf   Excited State Muon Transfer in \\
\vspace*{2mm}
 Hydrogen/Deuterium Mixtures} 
\end{center}

\begin{center}

\vspace*{5mm}

{\large

B. Lauss, P.Ackerbauer, W.H.Breunlich, B.Gartner,\\ 
M.Jeitler, $\rm P.Kammel^{*}$, J.Marton, W.Prymas, J.Zmeskal\\
Institute for Medium Energy Physics,\\
Austrian Academy of Sciences,\\
Boltzmanngasse 3, A-1090 Wien, Austria

D.Chatellard, J.-P.Egger, E.Jeannet\\
Institut de Physique de l'Universit\'e, CH-2000 Neuch\^atel, 
Switzerland

H.Daniel, A.Kosak, F.J.Hartmann\\
Physik Department, TU M\"unchen, D-85748 Garching, Germany

C.Petitjean\\
Paul Scherrer Institut, CH-5232 Villigen, Switzerland
}
\end{center}

\vspace*{10mm}

\begin{quote}

\small

We report the first direct observation of excited state
muon transfer in hydrogen/deuterium mixtures by direct 
measurement of $\rm q_{1s}$, the probability that a $\rm \mu p$ atom, 
which is initially formed in an excited state, reaches the 1s ground
state. The dependence of 
$\rm q_{1s}$ on deuterium concentration $\rm c_{d}$ was measured 
for two different densities at cryogenic temperatures using
charge coupled devices (CCDs) to detect the muonic X rays. First 
results 
based on the analysis of the $\rm K_{\alpha}$-lines of the two 
isotopes are presented.

\end{quote}

\newpage

\normalsize

One of the longstanding problems regarding the muonic cascade and 
muon catalyzed fusion ($\rm \mu CF$) in mixtures of hydrogen isotopes 
is the 
poor understanding of excited state transfer of the muon. 
Excited state transfer largely influences
the initial population of muonic atoms in the ground state, the 
starting
point for a complex sequence of processes leading to muon catalyzed 
fusion
\cite{Annu,Zme90,Kammel90,Acker93a}.

After a free muon is injected into a 
hydrogen (H), deuterium (D) or tritium (T) target, it is slowed down,
and an excited muonic atom is formed \cite{Leon62,Coh83}.
Rapid deexcitation to the ground state follows via various
competing collisional and radiative processes \cite{Marku81}. 
In the first approach the fraction of muons reaching the 1s level 
of the 
lighter isotope was considered to be proportional 
to the atomic concentration \cite{Gerst80} until the importance
of the possibility of a fast excited state transfer e.g.
\begin{eqnarray}
\nonumber   
p \mu (nl) + d  \Longrightarrow  p + d \mu (nl') + 135/n^{2} eV
\end{eqnarray}
(p=proton, d=deuteron)
which can compete with the other cascade processes was pointed
out \cite{Pono84}.  
A strong dependence on density $\rm \Phi$, concentration $\rm c_{d}$,
and collision energy $\rm \epsilon$ was 
predicted for $\rm q_{1s}$. ($\rm q_{1s}$ denotes the 
probability that
a $\rm \mu p$ atom, which is initially formed in an excited state,
reaches the 1s ground state.) Several calculations of this
functional dependence $\rm q_{1s}(\Phi,c_{i})$ have been published
for the H/D, H/T and D/T mixtures in the last decade 
\cite{Pono84,Pono85,Czap94a,Marku94b,Froe,Froe2,Asch95}. 

Until now, no direct measurement of $\rm q_{1s}$ has been performed.
The only experimental results on 
$\rm q_{1s}$ so far have been obtained
by indirect methods (``cycling rate analysis'') in the case of
DT-fusion e.g. Ref.\cite{Acker93a}.

Therefore, we performed an 
experiment in H/D mixtures \cite{Prop91} using an entirely
new approach based on the observation of the 
$\rm \mu p$ and $\rm \mu d$ X-ray intensities.
The small energy difference 
and the very low energy of these muonic X rays 
(energy of $\rm K_{\alpha}^{p\mu}=1.9$ keV, 
$\rm K_{\alpha}^{d\mu}=2.0$ keV)
were the main challenges of this measurement.

This experiment took place at the high intensity muon beam of the 
$\rm \mu E4$ area at PSI (Paul Scherrer Institut, Villigen, 
Switzerland). 
The setup is shown in figure 1.
The incoming muon beam was defined by scintillation counters and an 
aluminum beam aperture with a radius of 35 mm for liquid 
(45 mm for gas)
measurements. A muon momentum of $\sim$38 MeV/c 
($\sim$29 MeV/c) was selected for the liquid (gas) runs.

Special silver-coated steel target cells have been developed
for this experiment. 
$\rm 12.5\mu m$ thick Kapton foils were used as target windows
for the measurements at liquid hydrogen density.   
For the gas measurements, $\rm 25\mu m$ thick Kapton windows
had to withstand pressures of up to 6 bar at temperatures around
30 K. The shape of the windows was optimized to allow
efficient detection of low energy X rays. 
The rest of the target cell was surrounded by
superinsulation to reduce radiation heating.
Another $\rm 12.5 \mu m$ thick Kapton window 
separated the vacuum vessel from the vacuum of the charge 
coupled devices 
(CCDs) in order to protect the detector.

A new gashandling system was built for the preparation of
the various H/D mixtures. The pressure and
temperature of the target were monitored continually. 
During the measurements,
various samples of the target content were extracted using a small 
capillary leading directly from the target to a quadrupole 
mass spectrometer
\cite{QMS} in order to determine the relative abundance 
of the two isotopes.

For the first time, CCDs  \cite{Var90a,Var90b} 
have been employed for the observation of the muonic 
hydrogen X rays. 
Due to their excellent background suppression 
\cite{Egger93a,Egger93b,Lauss93} they provide a unique mean 
for the detection of X rays in this low energy region.
The size of each of the two CCD chips used was 
$\rm \sim   25  x  17  mm^{2} $ \cite{EEV}. 
The main chip component
was silicon with small absorption layers of $\rm SiN_{3}$ and 
$\rm SiO_{2}$ on the surface \cite{EEV}.
The depletion region had a thickness of $\rm \sim 30\mu m$.
To shorten the read-out time, each CCD chip
was split into two electronically independent detection areas.

``Single pixel analysis'' was used 
for the separation of  ``true'' X-ray hits from charged particles
or cosmic background. A ``single pixel hit'' 
was considered to be a ``true'' X ray if the charge content of the
surrounding 8 neighbour pixels was statistically compatible with
the noise peak of the CCDs \cite{Lauss93,Sigg94}.

Systematic measurements were performed for various deuterium 
concentrations 
at two different target densities, namely at liquid 
hydrogen density 
\linebreak (LHD =$\rm 4.25x10^{22}$ 
$\rm atoms/cm^{3}$) 
and at $\rm 1\% $ of LHD with temperatures of $\rm 20 - 30 K$.
Figure 2 shows the observed energy spectra of muonic deuterium, 
muonic hydrogen, and
of an isotopic mixture at liquid hydrogen density.
The small peak visible at 1.74 keV is the electronic 
$\rm K_{\alpha}$-line of silicon due to fluorescence of 
the CCD-materials.
In the mixture with $\rm c_{d}=0.25$ the effect of excited state 
transfer
can clearly be seen. Without considering excited state transfer,
one would expect the intensity of the deuterium 
$\rm K_{\alpha}$-line
to be about half of the adjoining $\rm K_{\alpha}^{p\mu}$-line 
(taking into 
account a $\rm \sim 40 \%$ difference in overall detection 
efficiency).

In the ``standard model'' of the muonic cascade 
\cite{Leon62,Marku81,Marku94}
the contribution of non-radiative processes to the
ground state deexcitation is suggested to be only a few percent
\cite{Marku81,Leon71}. 
This allows to determine $\rm q_{1s}$ in H/D mixtures
by disentangling the intensities of the $\rm \mu p$ and
$\rm \mu d$ K-lines.
In this first analysis
we considered only $\rm K_{\alpha}$ intensities.
Furthermore, the atomic capture probability was considered 
to be equal for hydrogen and deuterium \cite{Gerst80}.
(Cohen et al. \cite{Coh83} calculated a difference of 
$\rm \sim 6 \%$.)
This approach yields
\begin{eqnarray}
\nonumber
  q_{1s}  \approx  \frac {K_{\alpha}^{p\mu}} {1-c_{d}},
\hspace*{10mm}  q_{1s}  \approx  \frac {1 - K_{\alpha}^{d\mu}} 
{1-c_{d}}
\end{eqnarray}
{\small ($\rm  
K_{\alpha}^{\it p\mu}$...Intensity of muonic hydrogen 2-1 
transition, 
$\rm K_{\alpha}^{\it d\mu}$...Intensity of muonic deuterium 2-1 
transition, 
$\it c_{d}$...deuterium concentration).}\\

The energy spectra were analysed using Gaussians to fit the peak areas.
While the peak positions were treated as free parameters,
the FWHM of the Gaussian 
distributions was constrained to obey the energy relation
\begin{eqnarray}
\nonumber
  \Delta E \hspace{2mm} FWHM [eV] = 2.355 * E_{c} * ( N^{2} +  
\frac {F*E} {E_{c}} )^{1/2}
\end{eqnarray}
{\small ($\rm \Delta$E FWHM ... full width at half maximum of 
the fitting
function [eV], 
2.355 ... r.m.s. - FWHM conversion factor,
$\rm E_{c}$ ... conversion energy for 
an electron-hole pair in silicon (3.68 eV), N ... r.m.s. 
transfer and readout noise of the CCDs 
($\rm \sim 13 e^{-} r.m.s $), F ... Fano factor ($\rm \sim 0.12$), 
E ... energy of the X ray [eV]) 
}\\
which is valid for Si-detectors \cite{Egger93a,Egger93b}. 
A $\rm \mu^{+}$ run and an empty target run proved that there were
no background peaks visible within the observed energy region.
Therefore the background function could be approximated by
the sum of a constant and a term depending 
linearly on energy.
The CERN MINUIT program package \cite{Minuit}
was employed for the fitting procedure.

An analysis of the obtained data requires the precise knowledge of
the energy dependence of the detector efficiency.
A Monte Carlo program \cite{Lauss92} was written 
to account for the various contributions to X-ray absorption
by target content, windows, and CCD materials.
The calculation of the intrinsic detection efficiency of 
the CCDs and 
the X-ray absorption of the Kapton foils 
was checked experimentally \cite{Pub96}.

In our measurement we clearly observed a change 
in the muonic X-ray intensity pattern of the hydrogen 
and deuterium 
cascade in isotopic mixtures at various deuterium 
concentrations and
densities.
This can be directly attributed to the 
existence of excited state transfer during the muonic cascade. 
Our results for the $\rm q_{1s}$ values at different densities
are displayed as a function of deuterium concentration in figure 3. 
Statistical errors and systematic errors due to uncertainties
in the efficiency of the setup contribute to the given error bars.
The data point at $\rm 20 \%$ deuterium concentration
is the result of a previous test experiment \cite{Lauss93,Acker93b}.
For liquid hydrogen density the solid line demonstrates the 
theoretical expectation of
Ref.\cite{Czap94a} for a collision energy 
$\rm \epsilon=1$ eV, 
the dotted line shows the calculation of Ref.\cite{Czap94a} 
for $\rm \epsilon=6$ eV, and
the dashed line was calculated in Ref.\cite{Marku94b}.

The data clearly reveal that the dependence of $\rm q_{1s}$ on
deuterium concentration is much weaker than predicted by 
standard cascade theory.
Various new ideas were published recently which 
offer alternative explanations and 
show better agreement
with our results at liquid hydrogen density.
Czaplinsky et al. \cite{Czap94a} assumed higher collision energies 
$\rm \epsilon$ of the muonic atoms.
New calculations of the muonic cascade in hydrogen 
\cite{Marku94} suggest that the energies of the muonic atoms
- influenced by various cascade processes -
are distributed in a complex way, especially 
Coulomb deexcitation processes have to be taken into account.
The resulting values for $\rm q_{1s}$ are given in fig.3 
(dashed-dotted line) \cite{Asch95}.
A new mechanism for deexcitation of the muonic atom via an excited
$\rm (dt\mu)^{*}$ molecule was proposed in \cite{Froe,Froe2} which
suggests an alternative non-radiative transition to the ground state.
The existence of this ``resonant sidepath'' 
in the $\rm \mu CF$-cycle 
- up to now
one calculation exists for D/T mixtures only - would 
influence significantly the theoretical expectation for 
$\rm q_{1s}$ and 
its dependence on $\rm c_{d}$.

Within error limits, our results agree with the ones obtained by
analysis of cycling rates in the case of 
D/T mixtures given in Ref.\cite{Acker93a}.

Figure 4 displays the density dependence of $\rm q_{1s}$ 
at different deuterium concentrations.
A large difference between data taken at liquid hydrogen 
density and at $\rm 1 \%$ of LHD can be seen.
A more detailed result on the density dependence of 
$\rm q_{1s}$ is expected from the final analysis, which will include 
measurements at $\rm 4 \%$ and  $\rm 8 \%$ of LHD.

Financial support by the Austrian Science Foundation, the Austrian
Academy of Sciences, the Swiss Academy of Sciences, the Swiss
National Science Foundation, Paul Scherrer Institute and the
German Federal Ministry of Research and Technology is gratefully
acknowledged. 
It is a pleasure to thank D.Sigg for his software support. 
We are grateful to E.Steininger and H.Weiss for
their help during the experiment.

\newpage

\begin{figure}[htb]
\centerline{\psfig{file=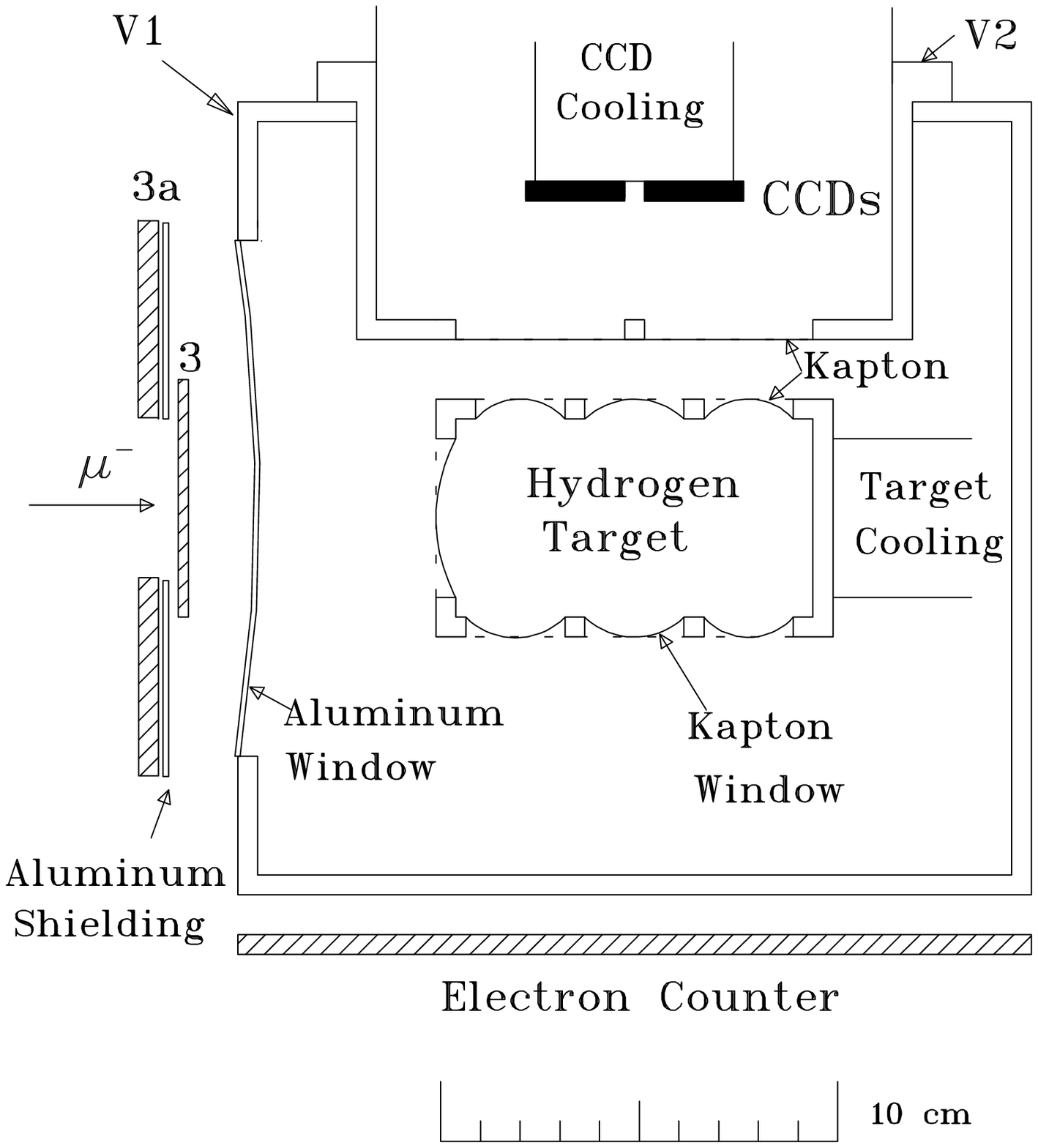,height=10.cm}}
\caption{
Top view of the setup of our experiment for the measurements in 
gaseous H/D mixtures. Two separated vacuum vessels were used 
for the target (V1) and the CCDs (V2). The scintillation counters 
3 and 3a were used to define incoming muons;
the electron counter served to detect electrons following 
muon decay.
For the measurements at liquid hydrogen density a smaller 
target cell 
( 50 x 50 x 34 mm ) was used.
}
\label{FIG.1}
\end{figure}

\begin{figure}[htb]
\centerline{\psfig{file=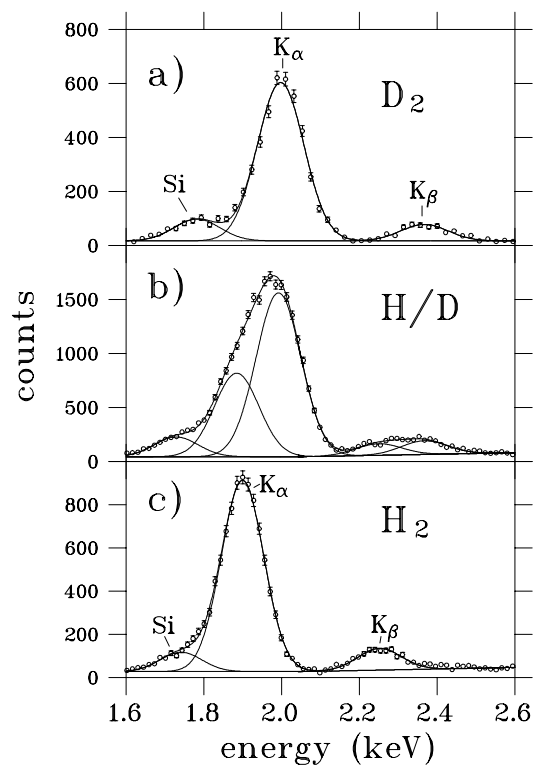,height=10.cm}}
\caption{
Energy spectra of a) liquid $\rm D_{2}$, b) a liquid 
H/D mixture containing $\rm 25 \%$ deuterium and 
c) liquid $\rm H_{2}$. The solid lines indicate the 
fits of the 
various peaks. 
}
\label{FIG.2}
\end{figure}

\newpage

\begin{figure}[htb]
\centerline{\psfig{file=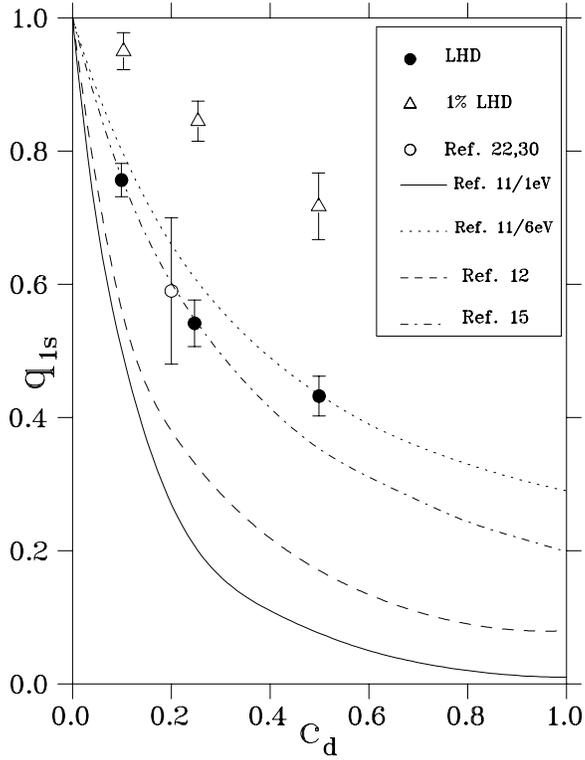,height=10.cm}}
\caption{
The dependence of $\rm q_{1s}$ on deuterium concentration. 
Filled dots display the results for liquid hydrogen density 
measurements,
the point at $\rm c_{d}=0.2$ is taken from Ref.
\cite{Lauss93,Acker93b}.
Triangles show the measurements at $\rm 1 \%$ of LHD. 
The theoretical expectations for $\rm q_{1s}$ at
liquid hydrogen density are taken from 
Ref.\cite{Czap94a} 
calculated for a collision energy
$\rm \epsilon=1$ eV (solid line),
$\rm \epsilon=6$ eV (dotted line),
Ref.\cite{Marku94b} (dashed line), and 
Ref. \cite{Asch95} (dashed-dotted line)
(scaling factor $\rm k_{t}$=0.5) . 
}
\label{FIG.3}
\end{figure}

\begin{figure}[htb]
\centerline{\psfig{file=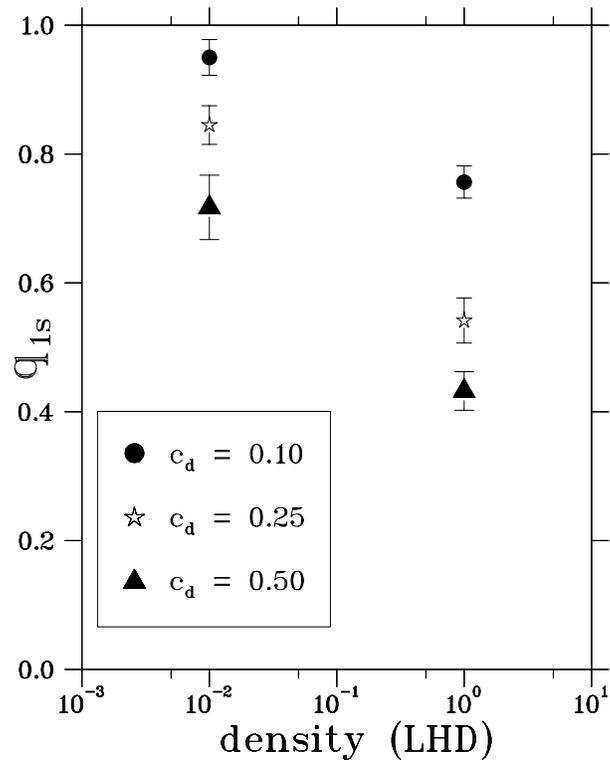,height=10.cm}}
\caption{
Dependence of $\rm q_{1s}$ on target density for three 
different deuterium concentrations. Dots refer to measurements 
at $\rm c_{d}=0.1$, stars to $\rm c_{d}=0.25$ 
and triangles to $\rm c_{d}=0.5$. The density is given relative 
to LHD. 
}
\label{FIG.4}
\end{figure}


\begin{thebibliography}{99}
\bibitem {Pub96} The results obtained in this measurements will be 
published elsewhere.
\end{thebibliography}
\end{document}